\documentclass[pra,aps,twocolumn,showpacs,epsfig]{revtex4}
\usepackage{color}
\usepackage{graphicx,amssymb,epsfig,epsf,amsmath}

\begin{document}

\title{Resonance States and Quantum Tunneling of Bose Einstein 
condensates in a 3D shallow trap}

\author{Sudip Kumar Haldar$^{1}$, Barnali Chakrabarti$^{1}$, 
Tapan Kumar Das$^{2}$}

\affiliation{$^{1}$ Department of Physics, Lady Brabourne 
College, P-1/2 Suhrawardy Avenue, Calcutta-700017, 
 India\\
$^{2}$ Department of Physics, University of Calcutta, 92 A.P.C. 
Road, Calcutta-700009, India.} 

%\vskip 1cm
\begin{abstract}
A correlated quantum many-body method is applied to describe 
resonance states of atomic Bose-Einstein 
condensates (BEC) in a realistic shallow trap (as opposed to 
infinite traps commonly used). The realistic 
van der Waals interaction is adopted as the interatomic 
interaction. We calculate  experimentally 
measurable decay rates of the lowest quasi-bound state in the 
shallow trap. The most striking result is the 
observation of a new metastable branch besides the usual one 
for attractive BEC in a pure harmonic trap. As 
the particle number increases the new metastable branch appears, 
then gradually disappears and finally 
usual metastable branch 
(associated with the attractive BEC in a harmonic trap) appears, 
eventually leading to the collapse of the condensate. 
\vskip 5pt \noindent
\end{abstract}
%\vskip 1cm
\pacs{03.75.Hh, 31.15.Xj, 03.65.Ge, 03.75.Nt.} 
%\vskip 1cm

\maketitle
\section{Introduction}
%\hspace*{2cm}
In experiments with Bose-Einstein condensation, the evolution of 
atomic cloud and its instability strongly 
depends on the external confinement, which is usually chosen as 
either isotropic or anisotropic pure ({\it i.e.} of 
infinite extent) harmonic potential. But, in the actual 
experimental setup, the trap is of finite 
extent. During the last few years, attention has been shifted to 
shallow optical dipole 
traps~\cite{Stamper}. As the quadratic trapping potential 
takes the shallow 
Gaussian envelope form, the anharmonicity of the 
potential must be taken into account. BEC in such a shallow trap of 
finite width supports resonance states 
which are quasi-bound. In such an experimental trap, the decay 
mechanism of the condensate becomes an 
important issue, as the condensate can escape from the trapping 
potential by quantum tunneling through 
intermediate barriers, in addition to the usual collapse of 
attractive condensates. Several attempts have 
been made to calculate life time of quasi-bound states and to 
study the transition from a resonance to a 
bound state by solving the Gross-Pitaevskii equation~(GPE) 
using contact delta 
interaction~\cite{Moiseyev,Paul,Korsch,Carr,Adhikari}. 
The GPE is based on the mean-field description and 
ignores correlations in the many-body wave function. But near the 
criticality, the energy of the resonance 
state becomes very close to the barrier height and the condensate 
becomes highly correlated. Thus 
incorporation of interatomic correlation becomes important. 
Naturally the full quantum many-body treatment  
incorporating a realistic interatomic interaction is indeed necessary. 
In particular the present experiments 
consider only a finite number of atoms in the external trap. Such 
condensates are quantum depleted, which 
again deserve a full quantum many-body treatment. The external 
shallow potential can be modeled by a 
quadratic plus a quartic potential, viz., 
$V(r)=\frac{1}{2}m\omega^{2}r^{2}+\lambda r^{4}$, where $\lambda$ 
is the anharmonic parameter. In our earlier work in this direction, 
we investigated the ground state 
properties in such a trap. We also observed dramatic change in 
the stability factor $\frac{N_{cr}|a_{sc}|}{a_{ho}}$ 
(where $N_{cr}$ is the critical number of atoms beyond 
which the attractive condensate 
collapses) due to the anharmonicity~\cite{Pankaj}.\\

%\hspace*{2cm}
Thus the aim of our present work is to employ a correlated quantum 
many-body approach incorporating a finite 
range realistic interatomic interaction to study resonance states 
in shallow traps. We calculate decay 
rates of quasi-bound condensates and study how the interatomic 
interaction and the anharmonicity interfere 
with  the decay process. Deviations from earlier works which used 
the mean-field equation 
can be attributed to two-body correlations and finite range of the 
realistic interaction. Due to the 
realistic nature of our calculation we expect results which are 
relevant to experiments. We observe 
resonance states associated with an unusual metastability for the 
repulsive BEC. The most striking result 
is the observation of two metastable branches for an attractive BEC. 
The attractive BEC containing $N$ 
atoms in a pure harmonic trap is associated with a metastable 
branch, which ultimately collapses for $N>N_{cr}$. In the 
present study we observe that with increase in particle number, 
a new metastability first 
appears, then gradually disappears and finally the usual 
metastability (associated with attractive BEC in a 
pure harmonic trap)  appears, eventually leading to the collapse. 
We also study macroscopic quantum 
tunneling and calculate decay rates for these two branches.\\

%\hspace*{2cm}

In Sec.~II we introduce the methodology which contains the 
many-body approach used in this work based on the 
correlated potential harmonic expansion method. Sec.~III contains 
numerical results and discussions. 
Conclusions are drawn following a summary of our work in Sec.~IV. 

\section{Correlated Potential Harmonic Expansion Method~(CPHEM)}
%\hspace*{2cm}
We adopt the potential harmonic expansion method with a short 
range correlation function~(CPHEM) which has 
already been established as a very successful and useful 
technique for the study of dilute BEC~\cite{Tapan,Das,Kundu}. 
Here we describe the technique briefly.\\

%\hspace*{2cm}
We consider that a system of $A=(N+1)$ identical bosons, 
each of mass $m$, is confined in an external trap 
($V_{trap}(r)$) which is modeled as a harmonic potential 
(of frequency $\omega$) perturbed by a quartic 
term. The time independent quantum many-body Schr\"odinger 
equation is written as:

\begin{eqnarray}
\Big[-\frac{\hbar^2}{2m}\sum_{i=1}^{A} \nabla_{i}^{2} 
+ \sum_{i=1}^{A} V_{trap}(\vec{x}_{i})& 
+\displaystyle{\sum_{i,j>i}^{A}} V(\vec{x}_{i}-\vec{x}_{j})\nonumber\\
-E\Big]\Psi(\vec{x}_{1},...,\vec{x}_{A})=0\hspace*{.1cm}\cdot
\end{eqnarray}
Where $E$ is the total energy of the system, 
$V(\vec{x}_{i}-\vec{x}_{j})$ is the two-body potential and 
$\vec{x}_{i}$ is the position vector of the $i$th particle. 
We define a set of $N$ Jacobi vectors as:
\begin{equation}
\vec{\zeta}_{i}=\sqrt{\frac{2i}{i+1}}(\vec{x}_{i+1}-
\frac{1}{i}\sum_{j=1}^{i} \vec{x}_j) \hspace*{.5cm}
 (i=1,...N)\cdot
\end{equation}
The center of mass coordinate is $\vec{R}= \frac{1}{A}
\displaystyle{\sum_{i=1}^{A}} \vec{x}_{i}$. As the 
labelling of particles is arbitrary, we choose the relative 
separation of $(ij)$-interacting pair 
$(\vec{x}_{ij}=\vec{x}_{i}-\vec{x}_{j})$ as 
$\vec{\zeta}_{N}$. We define the hyperradius $r$ as~\cite{Ballot}
\begin{equation}
r^{2} = \sum_{i=1}^{N}\zeta_{i}^{2} = \frac{2}{A}
\sum_{i,j>i}^{A}x_{ij}^{2} = 2\sum_{i=1}^{A} r_{i}^{2}
\end{equation}
where $\vec{r}_{i}=\vec{x}_{i}-\vec{R}$ is the position vector 
of $i$-th particle with respect to the center
of mass of the system. The relative motion of the bosons is given by 
\begin{eqnarray}
\Big[-\frac{\hbar^{2}}{m}\sum_{i=1}^{N} 
\nabla_{\zeta_{i}}^{2}+V_{trap}& + &V_{int}
(\vec{\zeta}_{1}, ..., \vec{\zeta}_{N})\nonumber\\
-E_{R}\Big]\Psi(\vec{\zeta}_{1}, ..., \vec{\zeta}_{N})& = & 
0\hspace*{.1cm}, 
\end{eqnarray}
$V_{trap}$ is the effective external trapping potential and 
$V_{int}$ is the sum of all pair-wise 
interactions. $E_R$ is the relative energy of the system {\it i.e.} 
$E=E_R+\frac{3}{2}\hbar\omega$. The 
laboratory BEC is designed to be very dilute, so that the probability 
of three or more atoms to come within 
the range of interaction is negligible. This is done in laboratory 
experiments, in order 
to avoid molecule formation through three body collisions. Hence, 
when the $(ij)$ pair interacts, the rest 
of the atoms are far apart and are inert spectators.
Therefore, only two-body correlations in the many-body wave 
function and 
pair-wise interactions among atoms
are important. For the $(ij)$-interacting pair, we define a hyperradius 
$\rho_{ij}$ for the remaining $(N-1)$
 noninteracting bosons as\cite{Fabre}
\begin{equation}
 \rho_{ij} = \sqrt{\sum_{K=1}^{N-1}\zeta_{K}^{2}}
\end{equation}
so that $r^2=x_{ij}^2+\rho_{ij}^2$. A hyperangle $\phi$ is introduced 
through $x_{ij}= r \cos\phi$ and $\rho_{ij}=r \sin\phi$. The full 
set of $3N$ hyperspherical variables are chosen as:

a) $r, \phi, \vartheta, \varphi$ ($\vartheta$ and $\varphi$ are 
the polar angles of $\vec{x}_{ij}$  corresponding to the 
interacting pair) and,

b) $(3N-4)$ hyperangular variables associated with the 
remaining $(N-1)$ inert spectators. 
Out of these, $2(N-1)$ are the polar angles 
associated with $(N-1)$ Jacobi vectors and another $(N-2)$ 
`hyperangles' define their relative lengths. 

In the Potential Harmonic Expansion Method~(PHEM), we decompose 
the many-body 
wave function($\Psi$) into Faddeev components $\phi_{ij}$ for the 
$(ij)$-interacting pair as

\begin{equation}
\Psi=\sum_{i,j>i}^{A}\phi_{ij}(\vec{x}_{ij},r)\hspace*{.1cm}\cdot
\end{equation}
Then Eq.~(4) can be expressed as
\begin{equation}
\left[T+V_{trap}-E_R\right]\phi_{ij}
=-V(\vec{x}_{ij})\sum_{kl>k}^{A}\phi_{kl}
\end{equation}
where $T = -\frac{\hbar^2}{m} \displaystyle{\sum_{i=1}^{N}} 
\nabla_{\zeta_{i}}^{2}$.
Since only two-body correlations are relevant, the Faddeev component 
$\phi_{ij}$ is independent of the coordinates of all the 
particles other than the interacting pair. 
Hence the total hyperangular momentum quantum number as also the orbital 
angular momentum of the whole system 
(comprising of all bosons) is contributed by the interacting pair only. 
We expand $\phi_{ij}$ 
in the subset of hyperspherical harmonics (HH) necessary for the 
expansion of $V(\vec{x}_{ij})$. Thus
\begin{equation}
\phi_{ij}(\vec{x}_{ij},r)
=r^{-(\frac{3N-1}{2})}\sum_{K}{\mathcal P}_{2K+l}^{lm}
(\Omega_{N}^{ij})u_{K}^{l}(r) \hspace*{.1cm}\cdot
\end{equation}
where $\Omega_{N}^{ij}$ denotes the full set of hyperangles in the 
$3N$-dimensional space corresponding to the $(ij)$-interacting pair. 
This new 
basis set $\{{\mathcal P}_{2K+l}^{lm}(\Omega_{N}^{ij})\}$ is called 
potential harmonics~(PH) 
basis and it is independent of $(\vec{\zeta}_{1}, ..., 
\vec{\zeta}_{N-1})$. Thus the total angular 
momentum of the condensate and its projection are simply $l$ and $m$, 
which comes from the 
interacting pair and all other quantum numbers coming from 
$(N-1)$ non-interacting bosons are 
kept frozen. An analytic form of the potential harmonics can 
be found in Ref~\cite{Fabre}. 
Substitution of Eq.~(8) into Eq.~(7) and taking projection 
on a particular PH gives~\cite{Tapan}
\begin{equation}
\begin{array}{ll}
&\Big[-\frac{\hbar^{2}}{m} \frac{d^{2}}{dr^{2}} +
\frac{\hbar^{2}}{m}\frac{{\mathcal L}_{K}({\mathcal L}_{K} 
+ 1)}{r^2} + V_{trap}(r) - E_{R} \Big]
U_{K}^{l}(r)\\
+&\displaystyle{\sum_{K^{\prime}}}f_{K^{\prime}l}^{2}
V_{KK^{\prime}}(r)
U_{K^{\prime}}^{l}(r) = 0
\hspace*{.1cm}\cdot
\end{array}
\end{equation}
when $V_{KK'}$ is the potential matrix and is given by
\begin{equation}
%\begin{array}{ll}
V_{KK^{\prime}}(r) =  
\int P_{2K+l}^{lm^*}(\Omega_{N}^{ij}) 
V\left(x_{ij}\right)
P_{2K^{\prime}+1}^{lm}(\Omega_{N}^{ij}) d\Omega_{N}^{ij} 
\hspace*{.1cm}\cdot
%\end{array}
\end{equation}
where ${\mathcal L}_{K} = 2K+l+\frac{3N-3}{2}$
and $f_{Kl}^2$ is the overlap of the PH for $ij$ partition with 
the sum 
of PHs of all partitions~\cite{Tapan}. $K$ is the grand 
orbital quantum number in $3N$ dimensional space. All other 
intermediate 
angular 
momentum quantum numbers take zero eigen values. As the number of 
active variables is  
now {\it only four} (global hyperradius $r$ and three other for  
$\vec{x}_{ij}$) 
for any number of bosons, the numerical complication is greatly 
simplified. Eq.~(9) 
can be put in symmetric form as
\begin{equation}
\begin{array}{cl}
&\Big[-\frac{\hbar^{2}}{m} \frac{d^{2}}{dr^{2}} +
V_{trap}(r) + \frac{\hbar^{2}}{mr^{2}}
\{ {\cal L}({\cal L}+1) \\
&+ 4K(K+\alpha+\beta+1)\}-E_R\Big]U_{Kl}(r)\\
+&\displaystyle{\sum_{K^{\prime}}}f_{Kl}V_{KK^{\prime}}(r)
f_{K^{\prime}l}
U_{K^{\prime}l}(r) = 0
\hspace*{.1cm},
\end{array}
\end{equation}\\
where ${\mathcal L}=l+\frac{3A-6}{2}$, $U_{Kl}=f_{Kl}u_{K}^{l}(r)$, 
$\alpha=\frac{3A-8}{2}$ and $\beta=l+1/2$.\\
%\hspace*{2cm}

In experimentally achieved BEC, as the energy of the interacting 
pair is 
extremely small, the two-body interaction is reproduced by the 
s-wave scattering 
length ($a_{sc}$). Positive (negative) $a_{sc}$ corresponds to a repulsive 
(attractive) condensate. 
In the Gross-Pitaevskii equation, the interatomic interaction is a 
contact interaction and is absolutely determined 
by its strength $a_{sc}$ only. Thus the two-body potential is purely repulsive 
or purely attractive, depending on the sign of $a_{sc}$. 
But van der Waals potential has two terms, 
one part represents a strong repulsion (usually represented by a 
hard core of radius $r_c$) at very short separation and the other 
part goes as $-\frac{C_6}{x_{ij}^{6}}$ and $a_{sc}$ can be either 
positive 
or negative depending on the value of $r_c$~\cite{Das}. Thus we 
determine $a_{sc}$ by solving the zero-energy 
two-body Schr\"odinger equation for the two-body wave function 
$\eta(x_{ij})$ 
\begin{equation}
-\frac{\hbar^2}{m}\frac{1}{x_{ij}^2}\frac{d}{dx_{ij}}\left(x_{ij}^2
\frac{d\eta(x_{ij})}{dx_{ij}}\right)+V(x_{ij})\eta(x_{ij})=0
\hspace*{.1cm}\cdot
\end{equation}
The value of $a_{sc}$ is determined from the asymptotic part of 
$\eta(x_{ij})$~\cite{Pethick}. The zero-energy two-body wave 
function 
$\eta(x_{ij})$ is also a good representation of the short range 
behavior of 
$\phi_{ij}$ and is taken as the two-body correlation function in the 
PH expansion basis, to improve the rate of convergence~\cite{CDLin}. 
Thus in the CPHEM, we replace Eq.~(8) by  
\begin{equation}
\phi_{ij}(\vec{x}_{ij},r)
=r^{-(\frac{3N-1}{2})}\sum_{K}{\mathcal P}_{2K+l}^{lm}
(\Omega_{N}^{ij})u_{K}^{l}(r) \eta(x_{ij}) \hspace*{.1cm}\cdot
\end{equation}
Introduction of $\eta(x_{ij})$ enhances the rate of convergence 
of the PH expansion dramatically. This has been actually 
verified in our numerical calculation. In our numerical procedure we 
solve Eq.~(12) for the zero energy two-body wavefunction $\eta(x_{ij})$ 
in the chosen two-body potential $V(x_{ij})$. We adjust the hard core 
radius $r_{c}$, such that $a_{sc}$ has the desired 
value~\cite{Das,Pethick}. 
This $\eta(x_{ij})$ is then used in Eq.~(13) and the potential 
matrix becomes
\begin{equation}
\begin{array}{cl}
&V_{KK^{\prime}}(r) =(h_{K}^{\alpha\beta} h_{K^{\prime}}^
{\alpha\beta})^{-\frac{1}{2}}\times \\
&\int_{-1}^{+1} \{P_{K}^{\alpha\beta}(z) 
V\left(r\sqrt{\frac{1+z}{2}}\right)
P_{K^{\prime}}^{\alpha \beta}(z)\eta\left(r\sqrt{\frac{1+z}{2}}\right)
W_{l}(z)\} dz \hspace*{.1cm}\cdot
\end{array}
\end{equation}
where $P_{K}^{\alpha\beta}(z)$ is the Jacobi polynomial, and its 
norm and 
weight function are $h_{K}^{\alpha\beta}$ and $W_{l}(z)$   
respectively~\cite{Abramowitz}. We truncate the $K$-sum in Rq.~(13) 
to an upper limit $K_{max}$ providing the desired 
convergence. Finally the coupled differential equation (CDE), 
Eq.~(11), is solved by the hyperspherical adiabatic approximation 
(HAA)~\cite{Coelho}. 
In HAA, one assumes that the hyperradial motion is slow compared 
to the 
hyperangular motion. Hence the latter is separated adiabatically 
and solved 
for a particular value of $r$, by diagonalizing the potential matrix 
together with the diagonal hypercentrifugal repulsion in Eq.~(11). 
The lowest eigenvalue, $\omega_0(r)$ (corresponding eigen column 
vector being $\chi_{K0}(r)$), provides the effective potential 
for the hyperradial motion.
We choose the lowest eigen 
potential ($\omega_{0}(r)$) as the effective potential in which 
the entire 
condensate moves as a single entity. Thus in HAA, the approximate 
solution 
(the energy and wave function) of the condensate is obtained by solving a 
single uncoupled differential equation
\begin{equation}
\left[-\frac{\hbar^{2}}{m}\frac{d^{2}}{dr^{2}}+\omega_{0}(r)-E_{R}
\right]\zeta_{0}(r)=0\hspace*{.1cm},
\end{equation}
subject to appropriate boundary conditions on $\zeta_{0}(r)$. 
The function $\zeta_{0}(r)$ is the collective wave function of 
the condensate in the hyperradial space. 
The lowest lying state in the effective potential $\omega_{0}(r)$ 
corresponds to the ground state of the condensate. The total energy 
of the condensate is obtained by adding the energy of the center of 
mass motion $(\frac{3}{2}\hbar\omega)$ to $E_{R}$.\\

%\hspace*{2cm}
The main advantages of our CPHEM are: 

i) Potential harmonic 
basis keeps all possible two-body correlations and yet the number 
of variables is reduced to only {\it four} for any number of bosons 
in the trap. So in spite of incorporating all the two-body 
correlations, we can treat quite a large number of atoms in the 
trap without much numerical complication. 

ii) We can use a realistic interatomic interaction like the van der 
Waals potential having a finite range, which itself takes care of 
the short range repulsion and interatomic correlations. 

iii) Unlike the GP 
equation, CPHEM does not have any pathological singularity, since 
the two-body interaction is a realistic one and has a strong 
short-range repulsion. 

Thus the 
CPHEM reveals the realistic picture. Clearly it is an improvement 
over the GP equation. Finally, by using the HAA, we reduce the 
multi-dimensional problem into an effective one-dimensional one 
in hyperradial 
space and the effective potential $\omega_{0}(r)$ of this 
one-dimensional  
problem provides a clear qualitative picture and a quantitative 
description of the system.\\
\vskip 1cm

\section{Results}

%\hspace*{2cm}
We choose the interatomic potential as the van der Waals potential 
with a hard core of radius $r_c$, {\it{viz.}}
$V(x_{ij})$= $\infty$  for  $x_{ij} \leq r_c$ and
= $-\frac{C_6}{x_{ij}^6}$  for  $x_{ij}>r_c$. 
The strong short range repulsion is parameterized by the hard 
core and the strength~($C_6$) is known for a given type of atom, 
{\it e.g.}, $C_{6}$ = $6.4898 \times 10^{-11}$ o.u. for Rb 
atoms~\cite{Pethick}. The value of $r_c$ is adjusted to get the 
desired value of $a_{sc}$. In oscillator unit~(o.u.) length and 
energy are given in the units of $a_{ho}=\sqrt{\frac{\hbar}
{m\omega}}$ and $\hbar\omega$ respectively. As $C_6 \rightarrow 0$, 
the potential becomes a hard core potential and $r_c$ coincides 
with the s-wave scattering length. As detailed in the previous 
section, we solve the zero-energy two-body Schr\"odinger 
equation for the interacting pair to get the 
value of $r_c$, which 
corresponds to the experimental scattering length $a_{sc}$.
With a tiny change in $r_{c}$, $a_{sc}$ may change 
by a large amount, including 
sign~\cite{Pethick}. With each additional change in sign, 
the potential supports an extra two-body 
bound state resulting in an additional node in $\eta(x_{ij})$. 
Thus the choice of 
$r_{c}$ is very crucial. We choose $r_c$ such that it 
corresponds to the zero-node in the two-body wave function 
for attractive interaction and one node for repulsive 
interaction~\cite{Kundu}.
For a repulsive BEC we choose $^{87}$Rb atoms with 
$a_{sc}$ = .00433 o.u. as in the JILA trap~\cite{Anderson}. For an 
attractive BEC, we choose $a_{sc} = -1.832 \times 10^{-4}$ o.u., 
which is one of the values as reported in the controlled 
collapse experiment of $^{85}$Rb atoms~\cite{Roberts,Cornish}. 
In both cases $r_c$ is determined by the method discussed above.\\ 

%\hspace*{2cm}
In the optical dipole trap, the trapping potential takes the 
shallow Gaussian form and the external trap is given by
$V(r)=\frac{1}{2}r^{2}+\lambda r^{4}$. For $\lambda > 0$ 
the frequency is blue shifted and for $\lambda < 0$ the frequency 
is red shifted. In the experiment~\cite{Stock,Bretin}, quartic 
confinement is created with a blue detuned Gaussian laser 
directed along the axial direction. The nonrotating condensate 
was cigar shaped and the strength of the quartic confinement 
was $\approx 10^{-3}$. In the present study we choose 
$\lambda$ as a controllable parameter and $|\lambda| \ll 1$. 
For $\lambda \textgreater 0$, as the quartic confinement becomes 
more tight, the frequency will increase for repulsive BEC and 
the attractive BEC will again be associated with a 
metastability~\cite{Pankaj}. These have been studied earlier 
both in one and three dimensions~\cite{Li,J}. However the 
most dramatic features 
are expected for $\lambda \textless 0$, and the potential can be 
easily approximated as $V\left(r\right) = 
\frac{1}{2}r^{2}$exp$\left(-cr^{2}\right)$ 
with $\lambda \approx \frac{c}{2}$. Our present 
calculation will consider only $\lambda \textless 0$.\\
\vskip 1cm
\subsection{Repulsive BEC}
%\hspace*{2cm}
For harmonic trapping with repulsive interaction, the condensate 
is always stable for any number of bosons. However due to the 
presence of anharmonicity we observe a new and different 
metastablity of the condensate. In Fig.~1 we plot the 
effective potential $\omega_0(r)$ as a function of $r$ for 
500 atoms of $^{87}$Rb in a shallow trap corresponding to 
$\lambda=-2 \times 10^{-5}$ o.u. and $a_{sc}=.00433$ o.u.. 
We observe a dramatic change in the effective 
\begin{figure}[hbpt]
%Fig.1
\vspace{-10pt}
\centerline{
\hspace{-3.3mm}
\rotatebox{0}{\epsfxsize=8.8cm\epsfbox{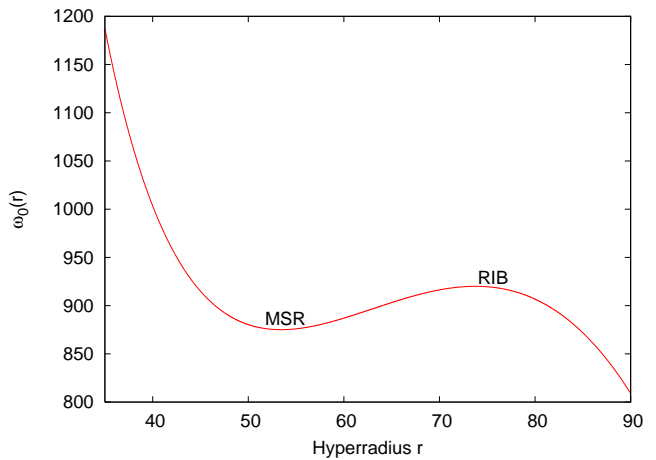}}}
\caption{(color online) Plot of the effective potential 
$\omega_0(r)$ against $r$ (both expressed in appropriate o.u.) 
for $N=500$ atoms of $^{87}$Rb with $a_{sc}$=.00433 o.u. 
and $\lambda=-2 \times 10^{-5}$ o.u. The metastable 
region (MSR) and the intermediate barrier on the right side (RIB) 
are indicated in the figure.}
\end{figure}
potential from that of a purely harmonic trap: a metastable region 
(MSR) appears bounded by an intermediate barrier on the right side 
(RIB), 
beyond which $\omega_{0}(r)$ decreases gradually, where the 
quartic term dominates over the quadratic term. In our earlier 
work~\cite{Pankaj}, we studied how the ground state properties 
and the low energy collective excitations get modified due to 
anharmonicity and calculated the stability factor 
$\frac{N_{cr}|a_{sc}|}{a_{ho}}$ in such a shallow trap. 
In the present work, we calculate the decay 
rate of quasi-bound states in the MSR, as the metastable condensate will 
tunnel through the intermediate barrier. The macroscopic 
tunneling rate is calculated semiclassically by the WKB 
tunneling formula:
\begin{eqnarray}
%\begin{equation}
\Gamma_{N}^{tunnel}& = &N\nu\exp(-2\int_{r_1}^{r_2}
\sqrt{2[\omega_0(r)-E]}\hspace*{.1cm}dr)\nonumber \\
                   & = &N\nu\exp(-2\sigma)\hspace*{.1cm}\cdot
%\end{equation}
\end{eqnarray}
where the limits of integration $r_1$ and $r_2$ are the inner 
and outer turning points of the intermediate barrier on the right~(RIB)  
of $\omega_0(r)$, $E$ is the energy 
of the metastable condensate 
and exp($-2\sigma$) is the WKB tunneling probability. The 
frequency of impact ($\nu$) of the condensate on the RIB 
is approximately given by
\begin{equation}
%\nu \sim \frac{1}{2 \int_{r_0}^{r_1} \frac{dr}
%{\sqrt{2\left[E-\omega_0(r)\right]}}}\hspace*{.1cm},
\nu \sim \Big[{2 \int_{r_0}^{r_1} \frac{dr}
{\sqrt{2\left[E-\omega_0(r)\right]}}}\Big]^{-1}\hspace*{.1cm},
\end{equation}
where $r_0$ and $r_1$ are the classical turning points of 
the metastable 
region. As $N$ increases, the net effect of the negative 
anharmonicity increases fairly rapidly. Hence, even though the minimum 
and stiffness of $\omega_0(r)$ increases with $N$, the 
difference ($\Delta \omega$) of the 
maximum of RIB ($\omega_{max}$) and the minumum of the 
MSR ($\omega_{min}$) decreases 
with increasing $N$. Consequently, RIB disappears 
($\Delta \omega=0$) when $N$ 
exceeds a critical value, $N_{cr}^{first}$ 
(to distinguish the critical numbers associated with the 
right side and the left side (see later) barriers, we 
name them as $N_{cr}^{first}$ and $N_{cr}^{second}$ 
respectively). This causes a new 
type of instability and eventual collapse. The tunneling rate is 
appreciable only when $E$ is close to $\omega_{max}$, and it 
increases rapidly as $E$ approaches $\omega_{max}$. 
In Fig.~2, the tunneling rate ($\Gamma_{N}^{tunnel}$) of 
the lowest resonance state is plotted against the number of 
condensate atoms close to the critical point for various 
values of anharmonic distortion. Near the criticality $N$ 
$\sim$ $N_{cr}^{first}$, the macroscopic tunneling is quite high and 
observation of this tunneling may be possible experimentally. 
The sharp peak near the criticality is attributed to the fact 
that the energy of the resonance state is close to the barrier 
height. Note that with increasing anharmonicity the 
right side barrier becomes lower, which makes $\Gamma_{N}^{tunnel}$ 
larger. For example, for $\lambda=-1.75 \times 10^{-5}$ o.u., the 
lowest resonance state near the critical point has tunneling 
probability 30\%, whereas for $\lambda=-2 \times 10^{-5}$ o.u. 
corresponding tunneling probability increases to 74\%. Consequently, 
$N_{cr}^{first}$ decreases with increasing $|\lambda|$.  
\begin{figure}[hbpt]
%Fig.2
\vspace{-10pt}
\centerline{
\hspace{-3.3mm}
\rotatebox{0}{\epsfxsize=9cm\epsfbox{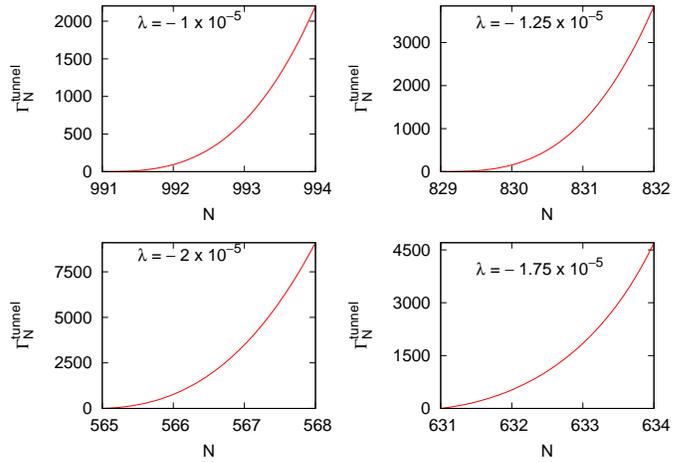}}}
\caption{(color online) Plot of $\Gamma_{N}^{tunnel}$  (in 
atoms per second) {\it vs} $N$ 
for the lowest resonance state near the criticality for 
various values of $\lambda$ (in o.u.).}
\end{figure}
In Fig.~3, we plot the resonance wave functions for two values 
of $N$ close to critical point ({\it viz.} $N=567$ and $N=568$, the 
critical number being $N_{cr}^{first} = 570$). The wave function within 
the metastable region is large and it has a small oscillatory part 
just outside the RIB. This 
clearly signifies that a part of the wave function leaks. For 
better clarity, the rapidly oscillatory part of the wave function 
is shown magnified in the inset of Fig.~3. Note that the amplitude 
of the leaked part increases as $N$ increases. 
At the critical point ($\Delta \omega=0$) the 
metastable region disappears and the whole wave function leaks, 
which corresponds to the collapse. The picture is qualitatively same 
as observed for attractive BEC in pure harmonic trap. However 
the phenomena near the present collapse is a bit different from the 
commonly observed collapse of attractive BEC in harmonic trap. 
In the latter case the metastable region is associated with a 
deep attractive well on the left side of the MSR, the metastable 
condensate tunnels through the left intermediate barrier (LIB) near 
the origin and settels down in the deep well to form clusters. 
In a typical attractive condensate
we have checked that the size of the well is $\sim 0.05\mu$m. 
Hence, due to the high two-body and three-body collision rates within 
such a narrow well, atoms form cluster. The width of this 
wavefunction in the narrow well is of the order of $0.003\mu$m, which 
is the order of the size of the atomic cluster. But in the 
present case the atoms which escape by tunneling outward 
will form a non-condensed Bose gas.\\
\begin{figure}[hbpt]
%Fig.3
\vspace{-10pt}
\centerline{
\hspace{-3.3mm}
\rotatebox{0}{\epsfxsize=8.8cm\epsfbox{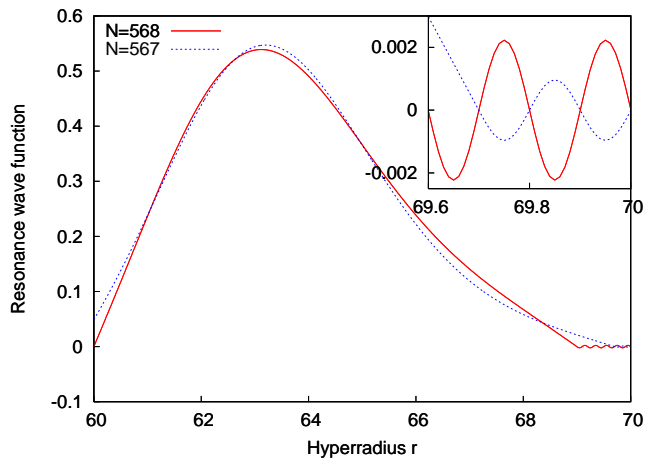}}}
\caption{(color online) Plot of the resonance wave function 
(in o.u.) {\it vs} r (in o.u.) 
for $N$=567 and $N$=568 for $\lambda=-2 \times 10^{-5}$ o.u. and 
$a_{sc} = .00433$ o.u.. The oscillatory part of the 
wavefunction immediately outside the barrier is shown magnified 
in the inset.}
\end{figure}

%\hspace*{2cm}
As a further study to observe transition from a resonance state 
to a bound state, we calculate decay rates for different 
values of effective interaction $|Na_{sc}|$. By decreasing 
the effective repulsive interaction, we effectively enhance attraction 
between the atoms. Table~1 clearly shows that even for very 
slow decrease in $|Na_{sc}|$, $\Gamma_N^{tunnel}$ decreases 
rapidly, and very soon reaches a vanishingly small value, which 
manifests the transition from resonance to a bound state.
\vspace*{.5cm}
\begin{center}
\hspace*{.5cm}Table 1: Decay rates of lowest resonance states for 
different $\lambda$ in a repulsive BEC ($a_{sc}=0.00433$ o.u.).\\
\vspace*{.3cm}
\begin{tabular}{|l|l|l|l|}
\hline
\multicolumn {2}{|c|}{$\lambda$ = $-1 \times 10^{-5}$ o.u.}
& \multicolumn{2}{|r|}{$\lambda$ = $-2 \times 10^{-5}$ o.u.}\\ \hline
$|Na_{sc}|$  &$\Gamma_{N}^{tunnel}$   &$|Na_{sc}|$   &$\Gamma_{N}^{tunnel}$\\ 
(o.u.) & (atoms/sec) & (o.u.) & (atoms/sec) \\ \hline 
4.30402    & 2202.6475               & 2.45944    & 9096.67862 \\ \hline
4.29969    & 53.7088                 & 2.45511    & 1738.7188 \\ \hline
4.29536    & 1.3083                  & 2.45078    & 108.5416   \\ \hline
4.29103    & 0.1107                  & 2.44645    & 1.5149 \\ \hline
4.28670    & 0.0158                  & 2.44212    & 0.1794 \\
\hline\hline
\end{tabular}
\end{center}
\vspace*{.3cm} 
Our result is qualitatively similar to earlier findings of Moiseyev 
{\it et.al}~\cite{Moiseyev}  where the transition from resonance to 
bound state was discussed. However the 
earlier calculations~\cite{Moiseyev,Carr} used singular delta function 
potential in the mean-field equation and a negative offset potential 
was required to facilitate conversion of a quasi-bound state into a 
bound state, in a three dimensional BEC. In our present calculation 
we need no such offset. 
This deviation from GP results is attributed to the use of a realistic 
interatomic interaction having a hard core repulsive part at shorter 
separation. Moreover, our results provide realistic aspects which are 
relevant to experiments.\\

\subsection{Attractive BEC}

%\hspace*{2cm}
The situation becomes more interesting for the attractive BEC in a 
shallow trap. We choose a condensate of $^{85}$Rb atoms 
with $a_{sc}$ = $-1.832 \times 10^{-4}$ o.u. 
For a clear understanding, we plot the effective 
potential in Fig.~4. The intermediate MSR is now bounded by two 
neighbouring barriers, one on the left (LIB) and one on the right (RIB) 
of unequal height. On the left side of LIB, a deep and narrow attractive 
well (NAW) appears. In the same 
vertical scale, we could not plot this deep well; hence it is not 
shown in Fig.~4. The RIB is the effect of negative anharmonicity 
which basically 
corresponds to a finite optical trap, whereas the LIB is the effect of 
the negative $a_{sc}$ which is commonly seen for attractive BEC in 
pure harmonic trap. The heights of the two barriers very strongly 
depend on two factors: first the anharmonic parameter and second 
the effective attractive interaction. Basically there is a competition 
between these two effects which causes the shape of the effective 
potential to change in a complicated fashion with the increase in $N$. 
So throughout our study we fix $\lambda = -9.37 \times 10^{-6}$ o.u.  
and the effective attractive interaction is tuned by changing the number 
of bosons. The metastable condensate will have a finite probability 
of macroscopic quantum tunneling through both the barriers.\\
\begin{figure}[hbpt]
%Fig.4
\vspace{-10pt}
\centerline{
\hspace{-3.3mm}
\rotatebox{0}{\epsfxsize=8.8cm\epsfbox{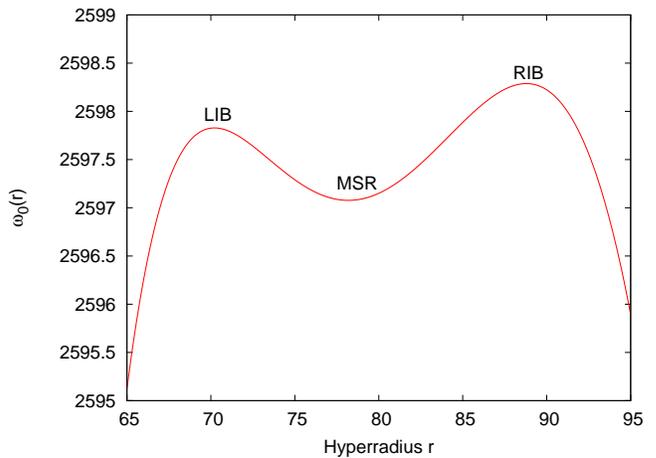}}}
\caption{(color online) Plot of the effective potential $\omega_{0}(r)$ in o.u. against 
hyperradius $r$ in o.u. for a condensate of $^{85}$Rb atoms with
$N=2660$, $\lambda=-9.37 \times 10^{-6}$ o.u. and 
$a_{sc} = -1.832 \times 10^{-4}$ o.u..}
\end{figure}
%\hspace*{2cm} 
We start with few hundreds $^{85}$Rb atoms in the trap, the two 
neighbouring barriers are quite high and tunneling of the condensate  
through either of them is negligible.  We have checked that in 
such a situation there is no substantial leakage of the condensate 
through the associated barriers. With further increase in particle 
number, we observe that the metastable region gradually becomes 
flatter, the corresponding condensate wave function expands slowly. 
With this wave function we calculate the average size of the 
condensate ($r_{av}$)~\cite{Anasua} as the root mean square distance 
of individual atoms from the center of mass of the condensate 
and is given by:
\begin{equation}
r_{av} = \Big<\frac{1}{A}\sum_{i=1}^{A}(\vec{x_{i}}-
\vec{R})^{2}\Big>^{1/2} = \frac{<r^2>^{1/2}}{\sqrt{2A}},
\end{equation}
where $\vec{R}$ is the center of mass coordinate. In fig.~5a, we 
plot $r_{av}$ as a function of $N$, for $N$ increasing from a 
few hundred to a few thousand bosons. We find that 
$r_{av}$ increases slowly as expected (as RIB decreases and LIB 
does not change substantially, and as a consequence, the wave 
function spreads outwards). Finally at $N=2460$, RIB vanishes 
and there is no MSR to hold the condensate, we call it a partial 
collapse. This is also reflected in the sharp fall in $r_{av}$ at 
$N=2460$ (Fig.~5a). Thus $N_{cr}^{first} =2460$. 
The associated tunneling rate $\Gamma_N^{tunnel}$ 
near the first collapse is shown in Fig.~6a. Near $N_{cr}^{first}$, 
the condensate is associated with a large tunneling probability. 
Thus $N_{cr}^{first}$ is associated with first branch of 
metastable condensate.\\
\begin{figure}[hbpt]
%Fig.5
\vspace{-10pt}
\centerline{
\hspace{-3.3cm}
\rotatebox{0}{\epsfxsize=9cm\epsfbox{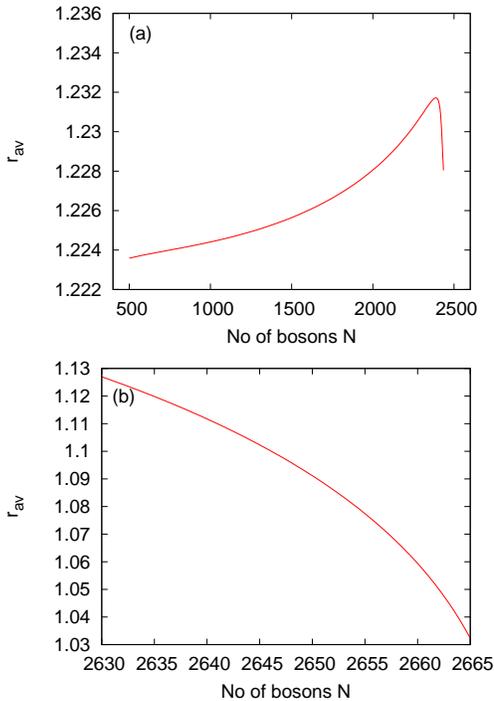}}}
\caption{(color online) Plot of average size of the attractive condensate 
$r_{av}$ (in o.u.) as a function of $N$ in the the anharmonic trap 
($\lambda = -9.37 \times 10^{-6}$ o.u. and 
$a_{sc} = -1.832 \times 10^{-4}$ o.u.) near the first [panel (a)] 
and second [panel (b)] criticality.}
\end{figure}
%\hspace*{2cm}
With further increase in particle number we observe that 
the MSR reappears at $N=2605$, the second branch starts to develop 
and LIB decreases gradually. This is due to the fact that the net 
attractive interaction now 
dominates over the effect arising from the anharmonic 
distortion, as the former increases as $\frac{N(N-1)}{2}$ 
while the latter increases as $N$. Due to substantial increase in 
attraction, both the height of LIB and the local minimum of 
$\omega_0(r)$ decrease 
rapidly, compared with the decrease of the height of RIB. Hence the 
MSR revives. 
As LIB decreases, the metastable 
condensate shrinks inwards, and we observe its behavior quite similar 
to what is seen in a 
pure harmonic trap: $r_{av}$ decreases sharply with increasing $N$, as 
seen in Fig.~5b. 
Unlike the first metastable branch, in the second branch the fall of 
$r_{av}$ is fairly sharp (note the difference in the horizontal scales 
in the two panels) and quicker collapse occurs at $N=2667$. 
We name this as the second criticality ($N_{cr}^{second}$). We also 
observe that 
near the second critical point, the condensate wave function is 
associated 
with an oscillatory  part in the left side of LIB. At 
$ N > N_{cr}^{second}$, the entire condensate collapses into the deep 
well, forming clusters. The associated tunneling rate 
$\Gamma_N^{tunnel}$ for the second metastable branch has been calculated
using Eq.~(16) and Eq.~(17) with the limits of the integrations suitably 
changed and is shown in Fig.~6b. 
\begin{figure}[hbpt]
%Fig.6
\vspace{-10pt}
\centerline{
\hspace{-3.3mm}
\rotatebox{0}{\epsfxsize=9cm\epsfbox{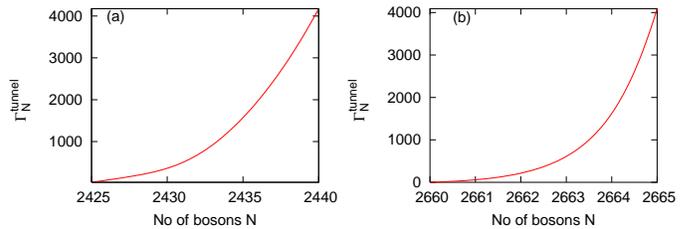}}}
\caption{(color online) Plot of $\Gamma_{N}^{tunnel}$ (in atoms/sec) against 
$N$ near the first [panel (a)] and the second [panel (b)] 
criticality.}
\end{figure}
The physical explanation for the appearance of two distinct 
metastabilities is as follows. The attractive interaction lowers the 
effective potential ($\omega_{0}(r)$) and the amount of lowering 
increases with $N$ as $\frac{N(N-1)}{2}$. This lowering decreases 
rapidly as $r$ increases. On the other hand, the anharmonic term also 
lowers the effective potential, but the corresponding lowering is 
appreciable only for large $r$ and it increases with $r$ as well as 
$N$. The increase with $N$ being roughly proportional to $N$. Hence, 
when $|\lambda|$ is not too small and $N$ increases from a small value 
($<N_{cr}^{first}$), the lowering due to anharmonicity (first 
lowering) at large $r$ is much larger than that due to the 
interaction (second 
lowering). Hence, with increasing $N$, $\omega_{0}(r)$ decreases for 
large $r$, giving rise to the first metastability and the appearance 
of the right 
intermediate barrier (RIB). As $N$ increases RIB decreases leading to 
the first criticality with the partial collapse at 
$N = N_{cr}^{first}$. 
With further increase of $N$, the second lowering at a smaller $r$ 
increases faster 
than the first lowering. This causes reappearance of the MSR, 
{\it whose minimum now gradually moves inwards}. As $N$ increases even 
further, the second lowering {\it for smaller r} increases very 
rapidly, inducing a deep attractive well and an intermediate barrier 
on the left (LIB) between this well and the MSR. The second lowering 
decreases very rapidly with increasing $r$ and is not strong enough 
at the position of RIB to alter it appreciably. As $N$ increases 
further, LIB disappears and the second criticality with collapse 
at $N = N_{cr}^{second}$ results. However the two branches are 
discontinuous in the range $2461 \leq N \leq 2604$ for the present 
choice of parameter sets. As we have said earlier, there is a 
competition between the two controllable parameters, {\it viz.} 
interaction 
and anharmonicity. Thus the existance of the discontinuous 
metastable branch will strongly depend on the choice of 
interaction and anharmonic distortion parameters. Our present 
study considers only a particular value of anharmonic distortion. 
So further study with other values of $\lambda$ is essential.\\

\section{Conclusion}
%\hspace*{2cm} 
In summary, we have applied a correlated many body method in 
three dimensions, incorporating a realistic interatomic interaction 
(van der 
Waals potential) to study metastable condensates confined in a 
trapping potential with a finite barrier. The potential is taken 
as a sum of a quadratic plus a quartic term, which approximates 
an optical dipole trap {\it i.e.} a harmonic confinement combined 
with a Gaussian envelope. We obtain the complete quantitative 
description  of the decay process of the quasi-stationary 
condensate with both repulsive and attractive interatomic 
interactions. Due to the 
use of a realistic interatomic interaction together with interatomic 
correlations in the many-body wave function and consideration 
of a finite number of atoms in the trap, 
our results exhibit more realistic picture. For a repulsive BEC, the 
quasi-stationary condensate can be stabilized by controlling the 
effective two-body interaction (through $a_{sc}$) and also the 
anharmonicity of the trap. By 
employing the WKB approximation, we calculate decay rates of such 
systems, which would be possible to measure experimentally. However, 
in contrast with earlier findings, in our present calculation no 
offset potential is required for the transition from a quasi-bound 
resonance state to a bound state. This difference is attributed to 
the use of a realistic interatomic interaction having a hard core 
at short range which prevents a catastrophic singularity at the 
origin as in the GP theory and produces a deep but finite well on 
the left of the left intermediate barrier. On the other hand, for 
an attractive BEC, in addition to the usual metastable condensate 
in a harmonic trap, we observe a new metastable branch which 
appears only for an intermediate range of particle number below the 
critical value for collapse due to attractive interaction only. The 
new metastable branch is also associated with an eventual collapse, 
for which the critical number is $N_{cr}^{first}$. We also determine 
the decay rates of the metastable BEC due to quantum tunneling from 
both the metastable regions. However, the transition between these 
two branches is discontinuous. We have observed that this 
discontinuity strongly depends on the distortion parameter. However 
as the decay rate of $^{85}$Rb atom in the new metastable region is 
quite high, the experimental study of this new phase may be difficult. 
But this technical difficulty may be circumvented by the 
proper choice of the parameters.\\

Prediction of two branches of criticality , in particular, the 
fact that the criticality associated with the right side barrier 
appears and then {\it disappears} as $N$ increases from a small value 
upto $N_{cr}^{first}$ and then beyond, are the most significant new physics outcome of 
this work. From a technical point of view, the use of a many-body 
theory, incorporating {\it all} two-body correlations in the 
many-body wave function and a {\it realistic} finite range 
interatomic interaction with a strong short-range 
repulsion are the realistic features. Deviations from earlier 
results are attributed to these. \\

\vskip 1cm
%\hspace*{2cm}
This work has been supported by a grant from the Department of Science 
and Technology (DST) [Fund No. SR/S2/CMP/0059(2007)], Government of 
India, under a research project. One of us (SKH) acknowledges the 
Council of Scientific and Industrial Research, India for the Junior 
Research Fellowship. TKD acknowledges the University Grants 
Commission, India for the Emeritus Fellowship.\\

\end{document}